\documentclass[a4paper,12pt]{article}
\usepackage{mathrsfs} \usepackage{amsmath} \usepackage{amssymb} \usepackage{slashed}
\usepackage{graphicx} \usepackage{caption}
\begin{document}
\newcommand{\ba}{\begin{eqnarray}} \newcommand{\ea}{\end{eqnarray}}
\newcommand{\be}{\begin{equation}} \newcommand{\ee}{\end{equation}}
\renewcommand{\figurename}{Figure}
\renewcommand{\thefootnote}{\fnsymbol{footnote}}

\vspace*{1cm}
\begin{center}
 {\Large\textbf{A Left-Right Mirror Symmetric Model: Common Origin of Neutrino Mass, Baryon Asymmetry and Dark Matter}}

 \vspace{1cm}
 \textbf{Wei-Min Yang}

 \vspace{0.3cm}
 \emph{Department of Modern Physics, University of Science and Technology of China}

 \emph{Hefei 230026, P. R. China}

 \emph{E-mail: wmyang@ustc.edu.cn}
\end{center}

\vspace{1cm}
\noindent\textbf{Abstract}:
  I suggest a left-right mirror symmetric particle model as the natural and aesthetic extension of the SM. As the left-right mirror symmetry breaking, the tiny neutrino mass is generated by the radiative mechanism, the baryon asymmetry through the leptogenesis arises from the characteristic decay of the TeV-scale mirror charged lepton, and a KeV-mass sterile Dirac fermion eventually becomes the CDM. The model can completely account for the common origin of the neutrino mass, the baryon asymmetry and the dark matter, moreover, profoundly uncover the internal connections among them. Finally, I discuss several feasible approaches to test the model predictions and probe the new physics by near future experiments.

\vspace{1cm}

\newpage
\noindent\textbf{I. Introduction}

\vspace{0.3cm}
  The standard model (SM) of the fundamental particles has successfully accounted for all kinds of the particle phenomena at or below the electroweak scale, refer to the relevant reviews in Particle Data Group \cite{1}. However, the SM has some shortcomings, aesthetically it is not a left-right symmetric framework, theoretically it can not at all address the three important issues of particle physics and cosmology: the tiny neutrino mass \cite{2}, the curious matter-antimatter asymmetry \cite{3}, and the mystical cold dark matter (CDM) \cite{4}. Up to now, particle and cosmology scientists have established plenty of experimental data of the neutrino physics and the baryon asymmetry \cite{1}, but the CDM has not yet been detected by any one terrestrial experiment except for the only evidence from the cosmic observations \cite{5}. The search for new physics evidences beyond the SM are always the focus of attention of the experimental physicists, there are worldwide a great deal of running and planed experiments which are aiming at the above goals. All of the investigations are gradually revealing the existence of an underlying and more fundamental theory beyond the SM.

  What style exactly is this new theory? A wide variety of the extensions of the SM have been suggested to address the above-mentioned issues in the last half century since the SM was established. A majority of them only focus on one of the issues and disregard possible connections among them, only a minority of them take into account an integrated solution, for example, the grand unified and supersymmetric theories, but these theories are either unbelievable complexity or unable to be tested. In any case, some inspiring and outstanding ideas are worth drawing lessons from. The tiny neutrino mass can be generated by the seesaw mechanism \cite{6} or arise from some loop-diagram radiative generation \cite{7}. The baryon asymmetry can be achieved by the thermal leptogenesis \cite{8} or the electroweak baryogenesis \cite{9}. The CDM candidates are possibly the sterile neutrino \cite{10}, the lightest supersymmetric particle \cite{11}, the axion \cite{12}, and so on. In recent years, some interesting models have exploited some connections among the neutrino mass, the baryon asymmetry and the CDM, for instance, the Scotogenic Model \cite{13}, the asymmetric CDM model \cite{14}, some sophisticated models \cite{15}, and the author's recent works on this field \cite{16}. Although progresses on new theory beyond the SM have been made all the time, a realistic and convincing theory is not established as yet. Therefore, finding out the correct new theory beyond the SM becomes the largest challenge for theoretical particle physics.

  Based on the universe harmony and the nature unification, it is very reasonable and believable that the tiny neutrino mass, the baryon asymmetry and the dark matter are related to each other and they have a common origin, in other words, a realistic theory beyond the SM should be able to unify the three things into a framework. On the other hand, this new theory should also abide by these two principles: the simplicity of the theory with fewer number of parameters, the feasibility of testing the theory by future experiments. If one theory is excessive complexity with too many parameters, then it is unbelievable, if it is unable to be tested, then it is also insignificant. After careful considerations and calculations, I here suggest a new left-right mirror symmetric model as the natural and aesthetic extension of the SM. This model with fewer number of parameters can completely account for the common origin of the above three things and profoundly uncover the internal connections among them. In addition, I give several feasible approaches to test the model and probe the new physics by means of the TeV-scale colliders, the neutrino experiments, the $\mu\rightarrow e\gamma$ LFV, and the high-energy cosmic rays.

  The remainder of this paper is organized as follows. In Section II, I outline the model and explain the neutrino mass generation. In Section III, I discuss the matter-antimatter asymmetry generation. The dark sector physics is discussed in Section IV. I give the numerical results and the model test approaches in Section V. Section VI is devoted to conclusions.

\vspace{0.6cm}
\noindent\textbf{II. Model and Neutrino Mass}

\vspace{0.3cm}
  The natural and aesthetic extension of the SM is implemented by introducing the mirror matter corresponding to the SM matter, meanwhile, the electroweak gauge group is extended to the left-right symmetric gauge group $SU(2)_{L}\otimes U(1)_{Y}\times SU(2)_{R}$, in addition, the two symmetries of a global $U(1)_{B-L}$ and a discrete $Z_{2}^{M}$ are sensibly imposed to constrain this extension.
\begin{table}
 \centering
\begin{tabular}{|c|c|c|}
 \hline &Left-handed Sector &Right-handed Sector\\ &(SM Matter) &(Mirror Matter)\\
 \hline Symmetry groups &\multicolumn{2}{|c|}{$SU(2)_{L}\otimes U(1)_{Y}\times SU(2)_{R}\otimes U(1)_{B-L}^{global}\otimes Z_{2}^{M}$}\\
 \hline Gauge fields &\multicolumn{2}{|c|}{\hspace{-0.5cm}$W_{L}^{\mu}(3,0,1)_{0}$\,,\hspace{0.5cm}$B^{\mu}(1,0,1)_{0}$\,,\hspace{0.5cm}$W_{R}^{\mu}(1,0,3)_{0}$}\\
 \hline Fermion fields
  &$q_{L}=\left(\begin{array}{c}u_{L}\\d_{L}\end{array}\right)(2,\frac{1}{3},1)_{\frac{1}{3}}$
  &$q_{R}=\left(\begin{array}{c}\xi_{R}\\\eta_{R}\end{array}\right)(1,\frac{1}{3},2)_{\frac{1}{3}}$\\
  &$d_{R}^{c}(1,\frac{2}{3},1)_{-\frac{1}{3}}$ &$\eta_{L}^{c}(1,\frac{2}{3},1)_{-\frac{1}{3}}$\\
  &$u_{R}^{c}(1,-\frac{4}{3},1)_{-\frac{1}{3}}$ &$\xi_{L}^{c}(1,-\frac{4}{3},1)_{-\frac{1}{3}}$\\
  &$l_{L}=\left(\begin{array}{c}\nu_{L}^{0}\\e_{L}^{-}\end{array}\right)(2,-1,1)_{-1}$
  &$l_{R}=\left(\begin{array}{c}\nu_{R}^{0}\\\chi_{R}^{-}\end{array}\right)(1,-1,2)_{-1}$\\
  &$e_{R}^{c}(1,2,1)_{1}$ &$\chi_{L}^{c}(1,2,1)_{1}$\\
  &$N_{L}^{0}(1,0,1)_{-1}$ &$N_{R}^{0}(1,0,1)_{-1}$\\
 \hline Scalar fields
  &$H_{L}=\left(\begin{array}{c}H_{L}^{0}\\H_{L}^{-}\end{array}\right)(2,-1,1)_{0}$
  &$H_{R}=\left(\begin{array}{c}H_{R}^{0}\\H_{R}^{-}\end{array}\right)(1,-1,2)_{0}$\\\cline{2-3}
  &\multicolumn{2}{|c|}{\hspace{-0.5cm}$\phi^{+}(1,2,1)_{2}$\,,\hspace{2cm} $\phi^{0}(1,0,1)_{0}$}\\
 \hline $Z_{2}^{M}$ parity &$+1$ &$-1$ but +1 for $W_{R}^{\mu}$ and $H_{R}$\\
 \hline
\end{tabular}
 \caption{The model particle contents and its symmetries. The color subgroup $SU(3)_{C}$ is omitted. The bracket following each field indicates its gauge quantum number and the right subscript of the bracket is its $B-L$ number, in addition, the right superscript of each component is its electric charge (but the quarks' charges are unlabelled). Note that $f_{R}^{c}=C\overline{f_{R}}^{T}$ is a left-handed anti-fermion and $f_{L}^{c}=C\overline{f_{L}}^{T}$ is a right-handed anti-fermion, where $C$ is the charge conjugation matrix. After the model symmetry breakings, $\xi,\eta$ and $\chi^{-}$ will become heavy-mass mirror quarks and charged lepton, which can however decay into the SM quarks and charged lepton. $\nu_{L}^{0}$ and $\nu_{R}^{0}$ will be combined into a tiny-mass Dirac neutrino through the loop-diagram radiative generation, which is the hot dark matter (HDM) in the present universe. $N_{L}^{0}$ and $N_{R}^{0}$ will be formed into a KeV-mass sterile and stable Dirac fermion, which is namely the CDM in the model. The four light neutral particles of $N^{0},\nu_{R}^{0},\phi^{0},\nu_{L}^{0}$, whose masses arise from the $Z_{2}^{M}$ breaking at the $\sim0.1$ MeV scale, successively decouple from the rest of the model particles and disappear into the dark sector, so they are difficult to be detected at the low energy.}
\end{table}
  Tab. 1 in detail shows the model particle contents and its symmetries, in which I omit the color subgroup $SU(3)_{C}$ since the strong interaction is not involved in the following discussions of this paper. The SM matter lie in the left-handed sector, while the mirror matter belong to the right-handed sector, explicitly, the model has the left-right mirror symmetry. Note that the local $U(1)_{Y}$ and the global $U(1)_{B-L}$ are common for the two sectors, so the three boson fields of $B^{\mu},\phi^{+},\phi^{0}$ are also common for the two sectors, in fact, their mirror particles are respectively themselves, in particular, $\phi^{0}$ is a real scalar field without any charges. The discrete $Z_{2}^{M}$ symmetry conserves a matter parity, under which the left-handed sector parity is ``$+1$", the right-handed sector parity is ``$-1$" but ``$+1$" for $W_{R}^{\mu}$ and $H_{R}$, note that $\phi^{+}$ has ``$+1$" parity and $\phi^{0}$ has ``$-1$" parity. The neutral singlet fermions $N_{L}^{0}$ and $N_{R}^{0}$ are respectively filled in the left-handed sector and the right-handed one, which will be combined into a Dirac fermion and eventually become the CDM in the model. More explanations are put in the caption of Tab. 1.

  All kinds of the chiral fermions in Tab. 1 have three generations as usual, they will obtain Dirac-type masses and form into Dirac fermions after the model symmetry breakings. There are no Majorana-type fermions in the model. By virtue of the fermion assignments and the explicit left-right mirror symmetry in Tab. 1, it is easily verified that all of the chiral anomalies are completely cancelled in the model, namely, the model is anomaly-free.

  We can now write the invariant Lagrangian of the model which satisfies the above-mentioned symmetries, it is composed of the three parts of the gauge kinetic energy terms, the Yukawa couplings and the scalar potentials. The gauge kinetic energy terms are
\begin{alignat}{1}
 \mathscr{L}_{\mathrm{G}}=&\:\mathscr{L}_{\mathrm{pure\:gauge}}
  +\sum\limits_{f_{L}}i\,\overline{f_{L}}\gamma_{\mu}D^{\mu}f_{L}+\sum\limits_{f_{R}}i\,\overline{f_{R}}\gamma_{\mu}D^{\mu}f_{R}\nonumber\\
 &+(D_{\mu}H_{L})^{\dagger}D^{\mu}H_{L}+(D_{\mu}H_{R})^{\dagger}D^{\mu}H_{R}+(D_{\mu}\phi^{+})^{\dagger}D^{\mu}\phi^{+}
  +\frac{1}{2}\partial_{\mu}\phi^{0}\partial^{\mu}\phi^{0},\nonumber\\
 D^{\mu}=&\:\partial^{\mu}+ig_{L}W_{L}^{\mu}\cdot\frac{\tau_{L}}{2}+ig_{Y}B^{\mu}\frac{Y}{2}+ig_{R}W_{R}^{\mu}\cdot\frac{\tau_{R}}{2}\,,
\end{alignat}
  where $f_{L}$ and $f_{R}$ denote all kinds of the chiral fermions in Tab. 1. $g_{L},g_{Y},g_{R}$  are three gauge coupling coefficients associated with the model gauge groups. $\tau_{i}$ are the three Pauli matrices and $Y$ is the charge operator of $U(1)_{Y}$.

  The Yukawa couplings are
\begin{alignat}{1}
 \mathscr{L}_{Y}=
 &\:q_{L}^{T}Y_{u}u_{R}^{c}H_{L}^{*}+q_{L}^{T}Y_{d}\,d_{R}^{c}\,\epsilon H_{L}
  +l_{L}^{T}Y_{e}\,e_{R}^{c}\,\epsilon H_{L}+\frac{1}{2}\,l_{L}^{T}Y_{L}\epsilon\,l_{L}\phi^{+}\nonumber\\
 &+q_{R}^{T}Y_{\xi}\,\xi_{L}^{c}H_{R}^{*}+q_{R}^{T}Y_{\eta}\eta_{L}^{c}\epsilon H_{R}
  +l_{R}^{T}Y_{\chi}\chi_{L}^{c}\epsilon H_{R}+\frac{1}{2}\,l_{R}^{T}Y_{R}\,\epsilon\,l_{R}\phi^{+}\nonumber\\
 &+\overline{u_{R}}\,Y_{1}\xi_{L}\phi^{0}+\overline{d_{R}}\,Y_{2}\eta_{L}\phi^{0}
  +\overline{e_{R}^{-}}\,Y_{3}\chi_{L}^{-}\phi^{0}+\overline{N_{L}^{0}}\,Y_{N}N_{R}^{0}\phi^{0}+\mathrm{H.c.}\,,
\end{alignat}
  where $\epsilon=i\tau_{2}$ is the two-order antisymmetric tensor. For concision I omit the charge conjugation matrix $C$ in those couplings of the first two lines, which is by default sandwiched between two spinors with the same chirality. The coupling parameters, $Y_{u},Y_{\xi},Y_{1}$, etc., are all $3\times3$ complex matrices in the flavor space, moreover, the leading matrix element of each coupling matrix should naturally be $\sim\mathcal{O}(1)$. Note that because of the spinor anti-commutativity and the $\epsilon$ antisymmetry, $Y_{L}$ and $Y_{R}$ must be two antisymmetric matrices for consistency. In Eq. (2), the $Z_{2}^{M}$ symmetry not only prohibits the explicit mass terms such as $\overline{e_{R}^{-}}M\chi_{L}^{-}, \overline{N_{L}^{0}}MN_{R}^{0}$, but also prevents the couplings such as $\overline{l_{L}}N_{R}^{0}H_{L}$, $\overline{l_{R}}N_{L}^{0}H_{R}$, $e_{R}^{-T}N_{R}^{0}\phi^{+}$, $\chi_{L}^{-T}N_{L}^{0}\phi^{+}$. Similarly, the $B-L$ conservation prohibits the terms such as $l_{L}^{T}N_{L}^{0}H_{L}^{*}$, $l_{R}^{T}N_{R}^{0}H_{R}^{*}$, $\overline{e_{R}^{-}}N_{L}^{0}\phi^{-}$, $\overline{\chi_{L}^{-}}N_{R}^{0}\phi^{-}$. Therefore the Yukawa couplings are greatly constrained by the model symmetries. Note that $N_{L}^{0}$ and $N_{R}^{0}$ can not couple to any gauge fields since they are both gauge singlets, on the other hand, they have no couplings to the other fermions owing to the $B-L$ and $Z_{2}^{M}$ symmetries, so $\overline{N_{L}^{0}}\,Y_{N}N_{R}^{0}\phi^{0}$ is the only coupling permitted for them. This thus leads that $N^{0}$ is naturally a sterile and stable fermion, eventually, it will become the CDM in the model. After the model symmetry breakings, the relevant scalar fields will develop their non-vanishing vacuum expectation values, as a result, Eq. (2) will give rise to all kinds of the fermion masses.

  Eqs. (1) and (2) explicitly show the left-right mirror symmetry which is defined as follows,
\begin{alignat}{1}
 &u_{L,R}\leftrightarrow\xi_{R,L},\hspace{0.3cm}d_{L,R}\leftrightarrow\eta_{R,L},\hspace{0.3cm}e_{L,R}^{-}\leftrightarrow\chi_{R,L}^{-},\hspace{0.3cm}
  \nu_{L}^{0}\leftrightarrow\nu_{R}^{0}\,,\hspace{0.3cm}N_{L}^{0}\leftrightarrow N_{R}^{0}\,,\nonumber\\
 &W_{L}^{\mu}\leftrightarrow W_{R}^{\mu}\,,\hspace{0.3cm} B^{\mu}\leftrightarrow B^{\mu},\hspace{0.3cm}
  H_{L}\leftrightarrow H_{R}\,,\hspace{0.3cm} \phi^{+}\leftrightarrow \phi^{+},\hspace{0.3cm} \phi^{0}\leftrightarrow \phi^{0},\nonumber\\
 &g_{L}=g_{R},\hspace{0.3cm} Y_{u}=Y_{\xi},\hspace{0.3cm} Y_{d}=Y_{\eta},\hspace{0.3cm} Y_{e}=Y_{\chi},\hspace{0.3cm} Y_{L}=Y_{R},\hspace{0.3cm}
  Y_{1,2,3}=Y_{1,2,3}^{\dagger}\,,\hspace{0.3cm} Y_{N}=Y_{N}^{\dagger}.
\end{alignat}
  This is indeed an aesthetics compared to the SM with many shortcomings. However, the exact left-right mirror symmetry can be relaxed by those equalities of the last line in Eq. (3) not being strictly valid.

  The full scalar potentials are
\begin{alignat}{1}
 V_{S}=&\:\mu_{L}^{2}H_{L}^{\dagger}H_{L}+\mu_{R}^{2}H_{R}^{\dagger}H_{R}+\mu_{+}^{2}\phi^{+}\phi^{-}+\frac{1}{2}\mu_{0}^{2}(\phi^{0})^{2}\nonumber\\
 &+\lambda_{L}(H_{L}^{\dagger}H_{L})^{2}+\lambda_{R}(H_{R}^{\dagger}H_{R})^{2}+\lambda_{+}(\phi^{+}\phi^{-})^{2}+\frac{1}{4}\lambda_{0}(\phi^{0})^{4}\nonumber\\
 &+2\lambda_{1}(H_{L}^{\dagger}H_{L})(H_{R}^{\dagger}H_{R})+[\lambda_{2}H_{L}^{\dagger}H_{L}+\lambda_{3}H_{R}^{\dagger}H_{R}](\phi^{0})^{2}\nonumber\\
 &+[2\lambda_{4}H_{L}^{\dagger}H_{L}+2\lambda_{5}H_{R}^{\dagger}H_{R}+\lambda_{6}(\phi^{0})^{2}]\phi^{+}\phi^{-}.
\end{alignat}
  The value areas of the mass-dimensional and dimensionless coupling parameters in Eq. (4) can completely control the vacuum configurations, and further determine the model symmetry breaking chain. It is natural and believable that the self-interaction of each scalar field is stronger but the interactions among them are weaker, therefore those interactive coupling parameters are much smaller than those self-coupling parameters in Eq. (4). In addition, we assume that the left-right mirror symmetry is explicitly broken by $|\mu_{L}^{2}|\ll|\mu_{R}^{2}|\sim(10^{6})^{2}\;\mathrm{GeV^{2}}$ in Eq. (4), which may arise from some symmetry breaking of a super-high scale physics, thus $H_{R}$ can first develop a non-zero vacuum expectation value at the $\sim10^{6}$ GeV scale, in later period $H_{L}$ and $\phi^{0}$ are successively induced to develop non-zero vacuum expectation values at the electroweak scale and the lower scale, but $\phi^{+}$ can not develop a non-zero vacuum expectation value, or it always keeps a vanishing vacuum expectation value. On the basis of an overall consideration, we therefore constrain all kinds of the parameters in Eq. (4) as follows,
\begin{alignat}{1}
 &(\lambda_{L},\lambda_{R},\lambda_{+},\lambda_{0})\sim10^{-1}>0\,,\hspace{0.3cm}
  10^{-6}\lesssim(|\lambda_{1}|,|\lambda_{2}|,\cdots,|\lambda_{6}|)\lesssim 10^{-2}\,,\nonumber\\
 &\mu^{2}_{R}\approx-\frac{v_{R}^{2}}{\lambda_{R}}\sim-(10^{6})^{2}\:\mathrm{GeV}^{2},\hspace{0.3cm}
  \mu^{2}_{L}<-\lambda_{1}v_{R}^{2}\,,\hspace{0.3cm} \mu^{2}_{0}<-\lambda_{3}v_{R}^{2}\,,\hspace{0.3cm} \mu^{2}_{+}>-\lambda_{5}v_{R}^{2}\,,
\end{alignat}
  where $\frac{v_{R}}{\sqrt{2}}=\langle H_{R}\rangle$ is the vacuum expectation value of the right-handed doublet scalar, see the following Eq. (6).

  Based on the limits of Eq. (5), we can directly derive the vacuum configurations from the $V_{S}$ minimum. The vacua of $H_{L}$ and $H_{R}$ are necessarily along the respective neutral component directions. The scalar sector will eventually appear three neutral and one charged scalar bosons under the unitary gauge. The detailed results are given as follows,
\begin{alignat}{1}
 &H_{L}\rightarrow\frac{h^{0}+v_{L}}{\sqrt{2}}\left(\begin{array}{c}1\\0\end{array}\right),\hspace{0.3cm}
  H_{R}\rightarrow\frac{\Phi^{0}+v_{R}}{\sqrt{2}}\left(\begin{array}{c}1\\0\end{array}\right),\hspace{0.3cm} \phi^{0}\rightarrow\rho^{0}+v_{0}\,,\hspace{0.3cm} \phi^{+}\rightarrow\phi^{+}\,,\nonumber\\
 &\left(\begin{array}{c}v_{0}^{2}\\v_{L}^{2}\\v_{R}^{2}\end{array}\right)=\left(\begin{array}{ccc}
  \lambda_{0}&\lambda_{2}&\lambda_{3}\\\lambda_{2}&\lambda_{L}&\lambda_{1}\\\lambda_{3}&\lambda_{1}&\lambda_{R}\end{array}\right)^{-1}
  \left(\begin{array}{c}-\mu^{2}_{0}\\-\mu^{2}_{L}\\-\mu^{2}_{R}\end{array}\right),\nonumber\\
 &v_{0}\sim0.1\:\mathrm{MeV}\ll v_{L}\approx246\:\mathrm{GeV}\ll v_{R}\sim10^{6}\:\mathrm{GeV},\nonumber\\
 &M_{h^{0}}\approx\sqrt{2\lambda_{L}}\,v_{L}\,,\hspace{0.3cm} M_{\Phi^{0}}\approx\sqrt{2\lambda_{R}}\,v_{R}\,,\hspace{0.3cm}
  m_{\rho^{0}}\approx\sqrt{2\lambda_{0}}\,v_{0}\,,\hspace{0.3cm} M_{\phi^{\pm}}\approx\sqrt{\mu^{2}_{+}+\lambda_{5}v_{R}^{2}}\,.
\end{alignat}
  $v_{R}$ is the breaking scale of $U(1)_{Y}\otimes SU(2)_{R}$, which is determined by the mirror sector physics. $v_{L}$ is namely the electroweak breaking scale, which has been fixed by the SM physics. $v_{0}$ is the $Z_{2}^{M}$ violating scale, which can be determined jointly by the neutrino mass and the dark sector physics. The mass-squared matrix of $(h^{0},\Phi^{0},\rho^{0})$ is approximately diagonal on account of the weaker couplings among the different scalars, so we can neglect the small mixings among them. In Eq. (6), $h^{0}$ is exactly identified as the SM Higgs boson with $M_{h^{0}}\approx125$ GeV. $M_{\Phi^{0}}$ is close to $v_{R}$, so the heavy $\Phi^{0}$ can not appear in the low-energy phenomena. $m_{\rho^{0}}$ is around $v_{0}$, so $\rho^{0}$ is a light dark scalar, which will play a role in the dark sector physics. $M_{\phi^{\pm}}$ is derived from the two contributions which are respectively the original mass $\mu_{+}$ and the induced mass from $\langle H_{R}\rangle$, however, its reasonable value should be $M_{\phi^{\pm}}\sim10^{4}$ GeV for the whole fit. $\phi^{\pm}$ will play a key role in generating the neutrino mass and the baryon asymmetry. In short, the limits of Eq. (5) are natural and reasonable from phenomenological point of view, they can ensure the vacuum stability and guarantee the following symmetry breaking chain.

  According to the assignments in Tab. 1 and the relations in Eq. (6), the model symmetries are spontaneously broken step by step through the following breaking chain,
\begin{alignat}{1}
 &SU(2)_{L}\otimes U(1)_{Y}\times SU(2)_{R}\otimes U(1)_{B-L}^{global}\otimes Z_{2}^{M}
  \xrightarrow{\langle H_{R}\rangle\sim10^{6}\:\mathrm{GeV}} \nonumber\\
 &SU(2)_{L}\otimes U(1)_{Y'}\otimes U(1)_{B-L}^{global}\otimes Z_{2}^{M}\xrightarrow{\langle H_{L}\rangle\sim10^{2}\:\mathrm{GeV}}\nonumber\\ &U(1)_{Q_{e}}\otimes U(1)_{B-L}^{global}\otimes Z_{2}^{M}\xrightarrow{\langle\phi^{0}\rangle\sim0.1\:\mathrm{MeV}}
  U(1)_{Q_{e}}\otimes U(1)_{B-L}^{global},\nonumber\\
 &Y'=Y+2I^{R}_{3},\hspace{0.5cm} Q_{e}=I^{L}_{3}+\frac{Y'}{2}=I^{L}_{3}+\frac{Y}{2}+I^{R}_{3},
\end{alignat}
  where $Y'$ is exactly identified as the SM hypercharge and $Q_{e}$ is namely the electric charge. This breaking chain is also aesthetical, the three hierarchical breaking transitions are very natural since they involve neither super-hierarchy nor super-high energy scale. Note that the global $B-L$ conservation is inviolate all the while, finally the residual gauge symmetry is only the local electric charge conservation. In addition, the $Z_{2}^{M}$ violation in fact occur in the dark sector, however, it is a surprising coincidence that its breaking scale, namely $\langle\phi^{0}\rangle=v_{0}\sim0.1$ MeV, is at the temperature of the nucleosynthesis onset in the SM sector. Although the spontaneous breaking of $Z_{2}^{M}$ can lead to the domain walls in the dark sector, the temperature fluctuation of the CMB produced by them is enough safely within the observation bound because $v_{0}$ is too low, namely $\frac{\delta T}{T}\sim10^{-5}\sqrt{\lambda_{0}}\,(\frac{v_{0}}{1\,\mathrm{MeV}})^{3}\ll10^{-5}$, refer to \cite{17}, therefore the domain walls have actually no effect on the universe evolution, we can ignore them in the model.

  As a result of the model symmetry breakings of Eqs. (6) and (7), all kinds of particle masses and mixings are generated through the Higgs mechanism. In the gauge sector, the masses and mixing of the gauge fields are given by the following relations,
\begin{alignat}{1}
 &D_{\mu}\rightarrow\partial_{\mu}+i\frac{g_{L}}{\sqrt{2}}(W_{L\mu}^{+}\tau_{L}^{+}+W_{L\mu}^{-}\tau_{L}^{-})
  +i\frac{g_{R}}{\sqrt{2}}(W_{R\mu}^{+}\tau_{R}^{+}+W_{R\mu}^{-}\tau_{R}^{-})\nonumber\\
 &\hspace{1.1cm}+ig_{L}Z_{L\mu}^{0}Q_{L}+ieA_{\mu}^{0}Q_{e}+ig_{R}Z_{R\mu}^{0}Q_{R},\nonumber\\
 &W_{L\mu}^{\pm}=\frac{W_{L\mu}^{1}\mp iW_{L\mu}^{2}}{\sqrt{2}}\,,\hspace{0.2cm}
  W_{R\mu}^{\pm}=\frac{W_{R\mu}^{1}\mp iW_{R\mu}^{2}}{\sqrt{2}}\,,\hspace{0.2cm}
  \left(\begin{array}{c}Z_{L\mu}^{0}\\A_{\mu}^{0}\\Z_{R\mu}^{0}\end{array}\right)=U_{13}U_{12}U_{23} \left(\begin{array}{c}W_{L\mu}^{3}\\B_{\mu}\\W_{R\mu}^{3}\end{array}\right),\nonumber\\
 &U_{23}=\left(\begin{array}{ccc}1&0&0\\0&c_{23}&s_{23}\\0&-s_{23}&c_{23}\end{array}\right),\hspace{0.2cm}
  U_{12}=\left(\begin{array}{ccc}c_{12}&-s_{12}&0\\s_{12}&c_{12}&0\\0&0&1\end{array}\right),\hspace{0.2cm}
  U_{13}=\left(\begin{array}{ccc}c_{13}&0&-s_{13}\\0&1&0\\s_{13}&0&c_{13}\end{array}\right),\nonumber\\
 &\tan\theta_{23}=\frac{g_{Y}}{g_{R}}\,,\hspace{0.3cm} \tan\theta_{12}=\frac{g_{R}}{g_{L}}\sin\theta_{23}\,,\hspace{0.3cm}
  \tan\theta_{13}\sim\frac{v_{L}^{2}}{v_{R}^{2}}\,,\nonumber\\
 &e=g_{L}\sin\theta_{12}\,,\hspace{0.3cm} Q_{e}=I^{L}_{3}+\frac{Y}{2}+I^{R}_{3},\nonumber\\
 &Q_{L}=\frac{I_{3}^{L}-Q_{e}\sin^{2}\theta_{12}}{\cos\theta_{12}}\,,\hspace{0.3cm}
  Q_{R}=\frac{I_{3}^{R}+(I_{3}^{L}-Q_{e})\sin^{2}\theta_{23}}{\cos\theta_{23}}\,,\nonumber\\
 &M_{W_{L}}=\frac{v_{L}g_{L}}{2}\,,\hspace{0.3cm} M_{Z_{L}}=\frac{M_{W_{L}}}{\cos\theta_{12}}\,,\hspace{0.3cm} m_{A}=0,\hspace{0.3cm}
  M_{Z_{R}}=\frac{M_{W_{R}}}{\cos\theta_{23}}\,,\hspace{0.3cm} M_{W_{R}}=\frac{v_{R}g_{R}}{2}\,,
\end{alignat}
  where $c_{ij}=\cos\theta_{ij}$ and $s_{ij}=\sin\theta_{ij}$ are mixing angles. It can be seen from the $Q_{L}$ expression that $\sin\theta_{12}$ is exactly identified as the Weinberg angle of the SM, so $\sin\theta_{12}=\sin\theta_{W}\approx0.481$ is actually fixed. If $g_{R}=g_{L}$, then $\sin\theta_{23}=\tan\theta_{12}=\tan\theta_{W}$ is also known. $\tan\theta_{13}$ is very small due to $v_{L}^{2}\ll v_{R}^{2}$, so we can ignore it. The masses of $W_{R}^{\pm}$ and $Z_{R}^{0}$ are $\sim10^{6}$ GeV, they will decay into the mirror quarks or leptons, but $Z_{R}^{0}$ can also decay into a pair of the SM quark or lepton in view of their $Q_{R}\neq0$, for example, $Z_{R}^{0}\rightarrow e^{-}+e^{+}$ and $Z_{R}^{0}\rightarrow\nu_{L}^{0}+\nu_{L}^{0c}$, these decays can thus become a source of high-energy cosmic rays. However, there are no $Z_{L}^{0}\rightarrow\nu_{R}^{0}+\nu_{R}^{0c}$ or $Z_{L}^{0}\rightarrow N^{0}+\overline{N^{0}}$ because both $\nu_{R}^{0}$ and $N^{0}$ have $Q_{L}=0$, this is of course consistent with the LEP bound on the invisible decay width of $Z_{L}^{0}$.

  In the Yukawa sector, the Yukawa couplings of Eq. (2) will undergo the following three steps of evolutions corresponding to the three steps of breakings in Eq. (7). After the first step breaking in Eq. (7), $\langle H_{R}\rangle$ gives rise to heavy masses of the mirror quarks and charged lepton, the Yukawa couplings thus evolve into
\begin{alignat}{1}
 \mathscr{L}_{Y}\xrightarrow{\langle H_{R}\rangle}
 &\:q_{L}^{T}Y_{u}u_{R}^{c}H_{L}^{*}+q_{L}^{T}Y_{d}\,d_{R}^{c}\,\epsilon H_{L}
  +l_{L}^{T}Y_{e}\,e_{R}^{c}\,\epsilon H_{L}+\frac{1}{2}\,l_{L}^{T}Y_{L}\epsilon\,l_{L}\phi^{+}\nonumber\\
 &-\overline{\xi_{L}}M_{\xi}\,\xi_{R}-\overline{\eta_{L}}M_{\eta}\eta_{R}-\overline{\chi_{L}^{-}}M_{\chi}\chi_{R}^{-}+\nu_{R}^{0T}Y_{R}\chi_{R}^{-}\phi^{+}\nonumber\\
 &+\overline{u_{R}}\,Y_{1}\xi_{L}\phi^{0}+\overline{d_{R}}\,Y_{2}\eta_{L}\phi^{0}
  +\overline{e_{R}^{-}}\,Y_{3}\chi_{L}^{-}\phi^{0}+\overline{N_{L}^{0}}\,Y_{N}N_{R}^{0}\phi^{0}+\mathrm{H.c.}\,,\nonumber\\
 &M_{\xi}=-\frac{v_{R}}{\sqrt{2}}Y_{\xi}^{T},\hspace{0.3cm}M_{\eta}=\frac{v_{R}}{\sqrt{2}}Y_{\eta}^{T},\hspace{0.3cm}M_{\chi}=\frac{v_{R}}{\sqrt{2}}Y_{\chi}^{T}.
\end{alignat}
  $M_{\chi},M_{\eta},M_{\xi}$ should be in the scope from several TeVs to hundreds of TeV. However, the mirror quarks and charged lepton can respectively decay into the SM quarks and charged lepton such as $\xi\rightarrow u+\phi^{0},\eta\rightarrow d+\phi^{0},\chi^{-}\rightarrow e^{-}+\phi^{0}$, and then the dark scalar $\phi^{0}$ can further decay into the CDM pair such as $\phi^{0}\rightarrow N^{0}+\overline{N^{0}}$, therefore the mirror quarks and charged lepton are completely decoupling and absence at the low-energy scale. Although they can not be detected at the present colliders, we can search their decay products through the high-energy cosmic rays. In addition, it can be seen from Eq. (9) that the annihilation processes such as $\chi^{\mp}+e^{\pm}\rightarrow N^{0}+\overline{N^{0}}$ via the s-channel $\phi^{0}$ mediation are gradually frozen as the universe temperature drops below $M_{\chi}$, thus $N^{0}$ will decouple from both the SM charged lepton and the mirror one, and then it disappears into the dark sector.

  The Yukawa couplings in Eq. (9) can inevitably generate an effective Dirac neutrino coupling by the loop-diagram radiative effect, this process is shown as Fig. 1.
\begin{figure}
 \centering
 \includegraphics[totalheight=4cm]{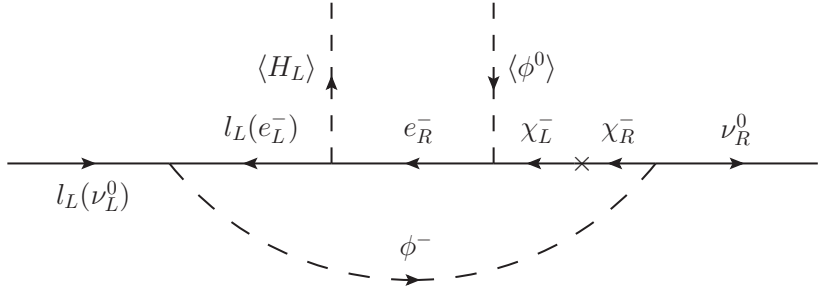}
 \caption{The loop diagram generating the effective Dirac neutrino coupling, by which the light neutrino mass is then achieved after developing $\langle H_{L}\rangle$ and $\langle\phi^{0}\rangle$ successively.}
\end{figure}
  Note that $\chi^{-}$ changes its chirality because of the $M_{\chi}$ insert in Fig. 1. The careful calculation gives the following results,
\begin{alignat}{1}
 &\mathscr{L}_{Neutrino}^{eff}=\frac{\sqrt{2}}{v_{L}}\,l_{L}^{T}Y_{\nu}\,\nu_{R}^{0c}H_{L}^{*}\phi^{0}+H.c.\,,\nonumber\\
 &(Y_{\nu})_{\alpha\beta}=\frac{v_{L}}{16\pi^{2}\sqrt{2}}\sum\limits_{i}(Y_{L}Y_{e}^{*}Y_{3})_{\alpha i}M_{\chi_{i}}(Y_{R}^{\dagger})_{i\beta}
  (C_{0}-p^{2}_{H_{L}}D_{11}+\slashed{p}_{\phi^{0}}\slashed{p}_{H_{L}}D_{12}-\slashed{p}_{\nu_{R}}\slashed{p}_{H_{L}}D_{13})\nonumber\\
 &\hspace{1.1cm}\approx\frac{1}{16\pi^{2}}\sum\limits_{i}\frac{(Y_{L}M_{e}^{\dagger}Y_{3})_{\alpha i}M_{\chi_{i}}(Y_{R}^{\dagger})_{i\beta}}
  {M^{2}_{\phi^{-}}}f(\frac{M^{2}_{\chi_{i}}}{M^{2}_{\phi^{-}}})\sim10^{-6}\,,\nonumber\\
 &C_{0}[(p_{l_{L}}-p_{H_{L}})^{2},p^{2}_{\nu_{R}},p^{2}_{\phi^{0}},m^{2}_{e},M^{2}_{\phi^{-}},M^{2}_{\chi_{i}}]
  =\frac{1}{M^{2}_{\phi^{-}}}f(\frac{M^{2}_{\chi_{i}}}{M^{2}_{\phi^{-}}})\,,\nonumber\\
 &f(\frac{M^{2}_{\chi_{i}}}{M^{2}_{\phi^{-}}})=\frac{ln\frac{M^{2}_{\chi_{i}}}{M^{2}_{\phi^{-}}}}{\frac{M^{2}_{\chi_{i}}}{M^{2}_{\phi^{-}}}-1}
  -\frac{i\,2\pi\Theta(\frac{M^{2}_{\chi_{i}}}{M^{2}_{\phi^{-}}}-1)}{\frac{M^{2}_{\chi_{i}}}{M^{2}_{\phi^{-}}}}\sim1,
\end{alignat}
  where $M_{\chi_{i}}(i=1,2,3)$ are the three mass eigenvalues of the mass matrix $M_{\chi}$ after it is diagonalized, and $M_{e}=\frac{v_{L}}{\sqrt{2}}Y_{e}^{T}$ is the charged lepton mass matrix (see the following Eq. (11)). $C_{0}$ and $D_{1i}$ are respectively the three-point and four-point functions of Passarino-Veltman \cite{18}. Because the $D_{1i}$ terms are much smaller than the $C_{0}$ term, for example, $p^{2}_{H_{L}}D_{11}\ll C_{0}$, we only calculate the $C_{0}$ term and ignore all of the $D_{1i}$ terms in the neutrino mass generation. $\Theta(x)$ is the step function, so Im$[C_{0}]=\frac{-i2\pi}{M^{2}_{\chi_{i}}}\neq 0$ only if $M^{2}_{\chi_{i}}>M^{2}_{\phi^{-}}$. Provided $M_{\chi_{1}}<M_{\chi_{2}}<M_{\chi_{3}}\sim M_{\phi^{-}}\sim10^{4}$ GeV and $M_{e}\sim1$ GeV, then we can estimate $Y_{\nu}\sim10^{-6}$, therefore, this effective neutrino coupling is very weak compared to those couplings in Eq. (9).

  The second step breaking in Eq. (7) is namely the electroweak breaking. $\langle H_{L}\rangle$ gives rise to the SM quark and charged lepton masses in Eq. (9), at the same time, the effective neutrino coupling in Eq. (10) develops into the normal neutrino coupling through $\langle H_{L}\rangle$, thus the Yukawa couplings of the fermions further evolve into
\begin{alignat}{1}
 \mathscr{L}_{Y}\xrightarrow{\langle H_{L}\rangle}
 &-\overline{u_{R}}M_{u}u_{L}-\overline{d_{R}}M_{d}\,d_{L}-\overline{e_{R}^{-}}M_{e}\,e_{L}^{-}+\nu_{L}^{0T}Y_{L}e_{L}^{-}\phi^{+}\nonumber\\
 &-\overline{\xi_{L}}M_{\xi}\,\xi_{R}-\overline{\eta_{L}}M_{\eta}\eta_{R}-\overline{\chi_{L}^{-}}M_{\chi}\chi_{R}^{-}+\nu_{R}^{0T}Y_{R}\chi_{R}^{-}\phi^{+}\nonumber\\
 &+\overline{u_{R}}\,Y_{1}\xi_{L}\phi^{0}+\overline{d_{R}}\,Y_{2}\eta_{L}\phi^{0}+\overline{e_{R}^{-}}\,Y_{3}\chi_{L}^{-}\phi^{0}
  +\overline{N_{L}^{0}}\,Y_{N}N_{R}^{0}\phi^{0}+\overline{\nu_{R}^{0}}\,Y_{\nu}^{T}\nu_{L}^{0}\phi^{0}+\mathrm{H.c.}\,,\nonumber\\
 &M_{u}=-\frac{v_{L}}{\sqrt{2}}Y_{u}^{T},\hspace{0.3cm}M_{d}=\frac{v_{L}}{\sqrt{2}}Y_{d}^{T},\hspace{0.3cm}M_{e}=\frac{v_{L}}{\sqrt{2}}Y_{e}^{T},
\end{alignat}
  where the normal neutrino coupling is naturally brought into $\mathscr{L}_{Y}$. Below the electroweak scale, the three light neutral particles $N^{0},\nu_{R}^{0},\phi^{0}$ are basically separated from the rest of the model particles, lastly $\nu_{L}^{0}$ will also decouple from the SM at the temperature of $\sim1$ MeV, thus all of them will eventually disappear into the dark sector.

  The last step breaking in Eq. (7) is that the $Z_{2}^{M}$ parity is violated by $\langle\phi^{0}\rangle=v_{0}\sim0.1$ MeV in the dark sector. As a result, this leads to light masses of $N^{0}$ and $\nu^{0}$, and also tiny mixings between the SM quark (charged lepton) and the mirror quark (charged lepton). Now we can completely obtain all of the fermion masses from Eq. (11), namely
\begin{alignat}{1}
 \mathscr{L}_{Y}\xrightarrow{\langle\phi^{0}\rangle}
 &-(\overline{u_{R}},\overline{\xi_{R}})\left(\begin{array}{cc}M_{u}&-v_{0}Y_{1}\\0&M_{\xi}^{\dagger}\end{array}\right)\left(\begin{array}{c}u_{L}\\\xi_{L}\end{array}\right)
  -(\overline{d_{R}},\overline{\eta_{R}})\left(\begin{array}{cc}M_{d}&-v_{0}Y_{2}\\0&M_{\eta}^{\dagger}\end{array}\right)\left(\begin{array}{c}d_{L}\\\eta_{L}\end{array}\right)\nonumber\\
 &-(\overline{e_{R}^{-}},\overline{\chi_{R}^{-}})\left(\begin{array}{cc}M_{e}&-v_{0}Y_{3}\\0&M_{\chi}^{\dagger}\end{array}\right)\left(\begin{array}{c}e_{L}^{-}\\\chi_{L}^{-}\end{array}\right)
  -\overline{N_{L}^{0}}M_{N}N_{R}^{0}-\overline{\nu_{R}^{0}}M_{\nu}\,\nu_{L}^{0}+\mathrm{H.c.}\,,\nonumber\\
 &M_{N}=-v_{0}Y_{N},\hspace{0.5cm} M_{\nu}=-v_{0}Y_{\nu}^{T}.
\end{alignat}
  Under the mass eigenstate basis, the three mass eigenvalues of $M_{N}$ are denoted by $m_{N_{i}}(i=1,2,3)$ and the ones of $M_{\nu}$ are denoted by $m_{\nu_{i}}(i=1,2,3)$. Obviously, the mixings between the SM quark (charged lepton) and the mirror quark (charged lepton) are very small because of $v_{0}\ll v_{L}\ll v_{R}$, so they can completely be neglected. There is no mixing between $N^{0}$ and $\nu^{0}$ by virtue of the model symmetry protection, therefore both $N^{0}$ and $\nu^{0}$ are stable particles without decay. $M_{N}$ is close to $v_{0}\sim0.1$ MeV, whereas $M_{\nu}$ is only $\sim0.1$ eV, so $N^{0}+\overline{N^{0}}$ can massively annihilate into $\nu^{0}+\overline{\nu^{0}}$ via the $\phi^{0}$ mediation but a tiny part of them is left over. In short, $N^{0}$ is authentically a sterile and stable WIMP, it eventually becomes the CDM, while $\nu^{0}$ becomes the HDM in the present universe. We will specially discuss the dark sector physics in Sec. IV.

  In Eq. (12), the neutrino mass matrix $M_{\nu}$ embraces the full information of the neutrino mass and mixing, which have mostly been measured by the experiments. Here we only work out the neutrino mass, regardless of its mixing. Provided that the Yukawa matrix equalities in Eq. (3) are valid, then $Y_{\nu}$ excluding the factor $f(\frac{M^{2}_{\phi^{-}}}{M^{2}_{\chi_{i}}})$ is a Hermitian matrix in Eq. (10), thus we can approximately derive the following results,
\begin{alignat}{1}
 &\mathrm{Tr}M_{\nu}=\sum\limits_{i}m_{\nu_{i}}=-\frac{v_{0}}{16\pi^{2}}
  \sum\limits_{i}\frac{(Y_{R}^{\dagger}Y_{L}M_{e}^{\dagger}Y_{3})_{ii}M_{\chi_{i}}}{M^{2}_{\phi^{-}}}f(\frac{M^{2}_{\chi_{i}}}{M^{2}_{\phi^{-}}})\nonumber\\
 \Longrightarrow &m_{\nu_{i}}\sim\frac{v_{0}\,m_{\tau}M_{\chi_{i}}}{16\pi^{2}M^{2}_{\phi^{-}}}|f(\frac{M^{2}_{\chi_{i}}}{M^{2}_{\phi^{-}}})|
  \lesssim10^{-10}\:\mathrm{GeV},
\end{alignat}
  where $m_{\tau}=1.777$ GeV is the largest eigenvalue of $M_{e}$, and we take $(Y_{R}^{\dagger}Y_{L}\frac{M_{e}^{\dagger}}{m_{\tau}}Y_{3})_{ii}\sim\mathcal{O}(1)$. In view of $M_{\chi_{3}}\sim M_{\phi^{-}}\sim10^{4}$ GeV, so $m_{\nu_{i}}$ is naturally Sub-eV. In conclusion, the model can naturally and successfully explain the tiny neutrino mass origin. Obviously, this mechanism is very different from a wide variety of seesaw ones \cite{19}.

  Based on the discussions in this Section, we finally summarize that the full particle mass spectrum in the model should be such relations as
\begin{alignat}{1}
 &m_{A}=0< m_{\nu_{i}}\lesssim0.05\:\mathrm{eV}\ll m_{N_{1}}\sim0.01\:\mathrm{MeV}< (m_{\rho^{0}},m_{N_{3}})\sim0.1\:\mathrm{MeV}\nonumber\\
 &< (M_{e},M_{q})\sim(10^{-3}-1)\:\mathrm{GeV}< (M_{W_{L}},M_{Z_{L}},M_{h^{0}},M_{t})\sim100\:\mathrm{GeV}\nonumber\\
 &< M_{\chi_{1}}\sim(1-10)\:\mathrm{TeV}< (M_{\phi^{-}},M_{\chi_{3}},M_{\eta},M_{\xi})\sim(10-10^{2})\:\mathrm{TeV}\nonumber\\
 &< (M_{W_{R}},M_{Z_{R}},M_{\Phi^{0}})\sim10^{3}\:\mathrm{TeV}.
\end{alignat}
  In the following Sections, we will also see that the mass relations of Eq. (14) can lead to successful explanations for the matter-antimatter asymmetry and the CDM.

\vspace{0.6cm}
\noindent\textbf{III. Baryon Asymmetry}

\vspace{0.3cm}
  In the model, the generation of the matter-antimatter asymmetry is directly associated with $\nu_{R}^{0}$ decoupling and disappearing into the dark sector, in fact, it arises from the following characteristic decays of the lightest mirror charged lepton $\chi_{1}^{\mp}$ which is the mirror particle of $e^{\mp}$. In the light of the Eq. (9) couplings and the Eq. (14) spectrum, $\chi_{1}^{-}$ with several GeV mass has only two decay modes at the tree level, i) the two-body decay of $\chi_{1}^{-}\rightarrow e^{-}+\phi^{0}$, which is dominant, ii) the three-body decay of $\chi_{1}^{-}\rightarrow\nu_{R}^{0c}+l_{L}+l_{L}$ via the heavier $\phi^{-}$ mediation, which is suppressed. When the effective neutrino coupling in Eq. (10) is taken into account, however, the three-body decay should also add a loop-diagram contribute. Fig. 2 draws the tree and loop diagrams of $\chi_{1}^{-}\rightarrow\nu_{R}^{0c}+l_{L}+l_{L}$ on the basis of the couplings in Eqs. (9) and (10).
\begin{figure}
 \centering
 \includegraphics[totalheight=4.5cm]{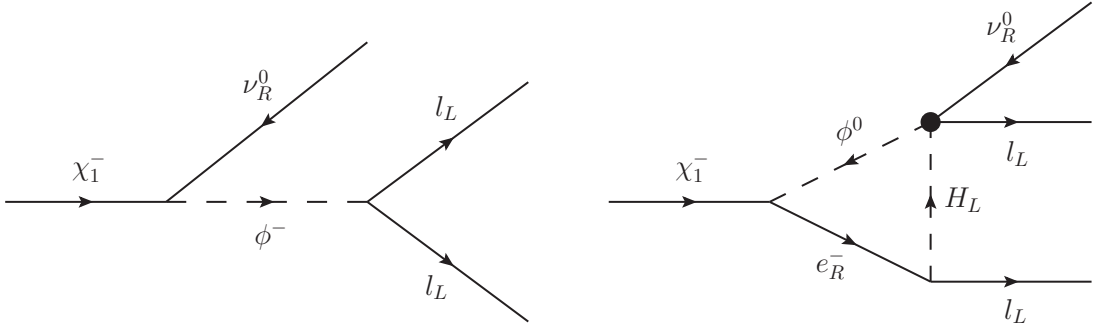}
 \caption{The tree and loop diagrams of the lightest mirror charged lepton three-body decay $\chi_{1}^{-}\rightarrow\nu_{R}^{0c}+l_{L}+l_{L}$. The heavy black vertex indicates the effective neutrino coupling. This is a $CP$-asymmetric and out-of-equilibrium decay but the $B-L$ number is conserved. The decay results in $\nu_{R}^{0}$ decoupling and disappearing into the dark sector. As a consequence, the $B-L$ asymmetry in the SM sector and the $-(B-L)$ asymmetry in the dark sector are simultaneously and equivalently generated, the latter is namely equal to the $\nu_{R}^{0}$ asymmetry, the former will be partly converted into the baryon asymmetry through the electroweak sphaleron effect.}
\end{figure}
  In addition, it should be emphasized that the $\chi_{1}^{-}$ decay processes completely conserve the $Y'$ charge, the $B-L$ number and the $Z_{2}^{M}$ parity, so they can only fulfil the other two of the Sakharov's three conditions \cite{20}.

  The decay process of Fig. 2 has the following three characteristics. The first characteristic is that the decay is a $CP$-asymmetric one, the decay rate of $\chi_{1}^{-}\rightarrow\nu_{R}^{0c}+l_{L}+l_{L}$ is different from one of its $CP$ conjugate process $\chi_{1}^{+}\rightarrow\nu_{R}^{0}+l_{L}^{c}+l_{L}^{c}$ due to the interference between the tree-diagram amplitude and the loop-diagram one. The decay Feynman amplitude is calculated as follows,
\begin{alignat}{1}
 &|\mathscr{M}|^{2}=\frac{m^{2}_{12}(M^{2}_{\chi_{1}}-m^{2}_{12})}{2M^{4}_{\phi^{-}}}\mathrm{Tr}[Y_{L}^{\dagger}Y_{L}](Y_{R}^{\dagger}Y_{R})_{11}
  -\frac{2m^{2}_{12}m^{2}_{23}}{M^{2}_{\phi^{-}}v_{L}^{2}}\mathrm{Re}[\mathrm{Tr}[Y_{\nu}Y^{\dagger}(M_{\chi_{1}})C_{12}^{*}]]\,,\nonumber\\
 &Y_{\nu}=\sum\limits_{i}Y(M_{\chi_{i}})C_{0}(M_{\chi_{i}})\,,\hspace{0.3cm}
  Y_{\alpha\beta}(M_{\chi_{i}})=\frac{1}{16\pi^{2}}(Y_{L}M_{e}^{\dagger}Y_{3})_{\alpha i}M_{\chi_{i}}(Y_{R}^{\dagger})_{i\beta}\,,\nonumber\\
 &\mathrm{Im}[C_{12}(m^{2}_{l},m^{2}_{23},M^{2}_{\chi_{1}},m^{2}_{e},M^{2}_{H_{L}},m^{2}_{\phi^{0}})]=\frac{-i\pi}{M^{2}_{\chi_{1}}}
  \left[1+\frac{M^{2}_{H_{L}}}{M^{2}_{\chi_{1}}}ln(1+\frac{M^{2}_{\chi_{1}}}{M^{2}_{H_{L}}})\right]\approx\frac{-i\pi}{M^{2}_{\chi_{1}}}\,,
\end{alignat}
  where $m^{2}_{12}=(p_{l}+p'_{l})^{2}$, $m^{2}_{23}=(p'_{l}+p_{\nu_{R}})^{2}$, and we specially define the functional matrix $Y(M_{\chi_{i}})$ to make concision of the expressions. In the Feynman amplitude, the first term is pure tree-diagram result, the second term is the $CP$-asymmetric interference term. Provided Eq. (3) being valid, then $Y(M_{\chi_{i}})$ is a Hermitian matrix, thus the factor of Tr$[Y(M_{\chi_{i}})Y^{\dagger}(M_{\chi_{1}})]$ is certainly real in the interference term, but the $C_{0}$ factor in $Y_{\nu}$ has the imaginary part of Im$[C_{0}(M_{\chi_{3}})]=\frac{-i2\pi}{M^{2}_{\chi_{3}}}\neq0$ in view of $M_{\chi_{3}}>M_{\phi^{-}}$, see Eq. (10), which becomes the only source of the $CP$-asymmetric decay, therefore, these two imaginary parts of Im$[C_{0}(M_{\chi_{3}})]$ and Im$[C_{12}]$ will jointly lead to a $CP$ asymmetry of the decay rate. In short, this mechanism of the $CP$ asymmetry generation does not completely depend on the $CP$-violating phases in the Yukawa matrices, it purely arises from the combined radiative effect of the two loop diagrams in Fig. 1 and Fig. 2. Finally, the relevant decay rates and $CP$ asymmetry are given as follows,
\begin{alignat}{1}
 &\Gamma[\chi_{1}^{-}\rightarrow e^{-}+\phi^{0}]=\frac{M_{\chi_{1}}}{32\pi}(Y_{3}^{\dagger}Y_{3})_{11}\,,\nonumber\\
 &\Gamma[\chi_{1}^{-}\rightarrow\nu_{R}^{0c}+l_{L}+l_{L}]=\frac{M_{\chi_{1}}}{768(2\pi)^{3}}(\frac{M_{\chi_{1}}}{M_{\phi^{-}}})^{4}
  \mathrm{Tr}[Y_{L}^{\dagger}Y_{L}](Y_{R}^{\dagger}Y_{R})_{11}\,,\nonumber\\
 &\Gamma_{total}[\chi_{1}^{-}]=\Gamma[\chi_{1}^{-}\rightarrow e^{-}+\phi^{0}]+\Gamma[\chi_{1}^{-}\rightarrow\nu_{R}^{0c}+l_{L}+l_{L}]
  \approx\Gamma[\chi_{1}^{-}\rightarrow e^{-}+\phi^{0}]\,,\nonumber\\
 &\varepsilon=\frac{\Gamma[\chi_{1}^{-}\rightarrow\nu_{R}^{0c}+l_{L}+l_{L}]
  -\Gamma[\chi_{1}^{+}\rightarrow\nu_{R}^{0}+l_{L}^{c}+l_{L}^{c}]}{\Gamma_{total}[\chi_{1}^{-}]}\nonumber\\
 &\hspace{0.2cm}=-\frac{M^{4}_{\chi_{1}}}{24v_{L}^{2}M^{2}_{\phi^{-}}(Y_{3}^{\dagger}Y_{3})_{11}}
  \mathrm{Tr}[\frac{Y(M_{\chi_{3}})Y^{\dagger}(M_{\chi_{1}})}{M^{2}_{\chi_{3}}M^{2}_{\chi_{1}}}]\nonumber\\
 &\hspace{0.2cm}\sim-\frac{m^{2}_{\tau}M^{3}_{\chi_{1}}}{24(16\pi^{2})^{2}v_{L}^{2}M^{2}_{\phi^{-}}M_{\chi_{3}}(Y_{3}^{\dagger}Y_{3})_{11}}\,,
\end{alignat}
  where the leading matrix elements of $Y_{L},Y_{R},Y_{3}$ are all $\sim\mathcal{O}(1)$ but these two elements of $(Y_{3}^{\dagger}Y_{3})_{11}$ and $(Y_{R}^{\dagger}Y_{R})_{11}$ are only $\sim10^{-6}$, and the trace of multiple matrix multiplication is also $\sim\mathcal{O}(1)$ in the last approximation. From Eq. (16), the three-body decay rate is $\sim10^{-7}$ times smaller than the two-body one because of the twofold suppressions of the phase space factor and the $(\frac{M_{\chi_{1}}}{M_{\phi^{-}}})^{4}$ factor. Provided $(Y_{3}^{\dagger}Y_{3})_{11}\sim10^{-6}$, $\frac{M_{\chi_{1}}}{M_{\phi^{-}}}\sim0.1$ and $\frac{M_{\chi_{3}}}{M_{\phi^{-}}}\gtrsim1$, then we can estimate $\varepsilon\sim 10^{-8}$, which is a suitable value for the successful leptogenesis.

  The second characteristic is that the decay is out-of-equilibrium, the three-body decay rate is smaller than the universe Hubble expansion rate, namely
\ba
 \Gamma[\chi_{1}^{-}\rightarrow\nu_{R}^{0c}+l_{L}+l_{L}]< H(T=M_{\chi_{1}})=\frac{1.66\sqrt{g_{*}}M^{2}_{\chi_{1}}}{M_{Pl}}\,,
\ea
  where $M_{Pl}=1.22\times10^{19}$ GeV, $g_{*}(T)$ is the effective number of relativistic degrees of freedom. At the temperature of $T=M_{\chi_{1}}$, the relativistic states include all the SM particles as well as the light dark particles $\phi^{0},N^{0},\nu_{R}^{0}$, so we can easily figure out $g_{*}=123.5$. Note that the two-body decay is still in equilibrium since $\Gamma[\chi_{1}^{-}\rightarrow e^{-}+\phi^{0}]\gg H(T=M_{\chi_{1}})$.

  The third characteristic is that the out-of-equilibrium decay directly results in $\nu_{R}^{0}$ decoupling from the rest of the model and disappearing into the dark sector. As a consequence of the above three characteristics, the Fig. 2 decay can simultaneously and equivalently generate a $B-L$ asymmetry in the SM sector and a $-(B-L)$ asymmetry in the dark sector which is namely equal to the $\nu_{R}^{0}$ asymmetry, but the total $B-L$ asymmetry is always vanishing in the whole universe. Note that $N^{0}$ can not be generated an asymmetry due to its unique coupling in the dark sector, so it is surely a symmetric CDM.

  After the $\chi^{\pm}$ decays are finished, all of the mirror particles including $\nu_{R}^{0}$ are completely decoupling. The generated $B-L$ asymmetry in the SM sector can be partly converted into the baryon asymmetry through the sphaleron process which is effectively put into effect above the electroweak scale \cite{21}. Therefore, the relevant asymmetries normalized to the entropy are given by the following relations \cite{17},
\begin{alignat}{1}
 &Y^{SM}_{B-L}=-Y^{DS}_{B-L}=Y_{\nu_{R}}=\frac{n_{\nu_{R}}-\overline{n}_{\nu_{R}}}{s}=\kappa\frac{-\varepsilon}{g_{*}}\,,\nonumber\\
 &Y_{B}=c_{s}Y^{SM}_{B-L},
\end{alignat}
  where $c_{s}=\frac{28}{79}$ is the sphaleron conversion coefficient in the SM sector, $s$ is the total entropy density in the SM and dark sectors, $\kappa$ is a dilution factor. In fact, we can take $\kappa\approx1$ because the dilution effect is almost vanishing as the universe temperature drops below $M_{\chi_{1}}$. In addition, it should be pointed out that the effective neutrino coupling in Eq. (10) is more severely out-of-equilibrium due to the suppression of $\frac{Y_{\nu}}{v_{L}}\sim10^{-8}\;\mathrm{GeV^{-1}}$, so it can not dilute these asymmetries in Eq. (18) at all. Below the electroweak breaking scale, the sphaleron process is closed, thereby the baryon asymmetry is locked down. In later time, the $\nu_{R}$ asymmetry and the $\nu_{L}$ one can partially be erased through the weak normal neutrino coupling appearing in Eq. (11), but this has no effect on the baryon asymmetry.

  As the universe temperature drops to the electroweak scale, then the universe comes into the SM epoch and the known evolutions, while the evolutions in the dark sector will specially be discussed in the next Section. In the present-day universe, the baryon asymmetry and its density are given by
\begin{alignat}{1}
  &\eta_{B}=\frac{n_{B}-\overline{n}_{B}}{n_{\gamma}}=\frac{s(T_{0})}{n_{\gamma}(T_{0})}Y_{B}\approx6.1\times10^{-10}\,,\nonumber\\
  &\Omega_{B}h^{2}=\frac{m_{p}\eta_{B}n_{\gamma}(T_{0})}{\rho_{c}}h^{2}\approx0.0223\,,
\end{alignat}
  where $T_{0}\approx2.73$ K is the present-day temperature of the CMB, $n_{\gamma}(T_{0})\approx411\:\text{cm}^{-3}$ is the photon number density, and $\frac{s(T_{0})}{n_{\gamma}(T_{0})}=3.6$ is because only the photon is still relativistic and the massive neutrino has become non-relativistic. $m_{p}=0.938$ GeV is the proton mass, $\rho_{c}=1.054\times10^{-5}h^{2}\:\text{GeV}\,\text{cm}^{-3}$ is the critical energy density \cite{1}. $\eta_{B}\approx6.1\times10^{-10}$ is the current baryon asymmetry measured by the multiple experiments \cite{1,22}, and $\Omega_{B}h^{2}\approx0.0223$ is the current baryon density \cite{1}. In conclusion, the model can clearly and successfully explain the origin of the baryon asymmetry through this novel leptogenesis mechanism, in particular, the matter-antimatter asymmetry is generated just at the TeV scale, so it is very possible to test this mechanism in the near future.

\vspace{0.6cm}
\noindent\textbf{IV. Dark Sector Physics}

\vspace{0.3cm}
  As the universe temperature decreasing, the three neutral particles of $N^{0},\nu_{R}^{0},\phi^{0}$ successively decouple from the hot plasma and disappear into the dark sector, but they can still interact with each other in the dark sector. Below the electroweak scale, there are two portals connecting the dark sector and the SM one, by which the SM sector can communicate with the dark one. One portal is the $\lambda_{2}$ coupling term in the Eq. (4) scalar potentials. Provided $10^{-5}\lesssim\lambda_{2}\lesssim10^{-3}$, the reaction rate of $h^{0}+h^{0}\leftrightarrow\phi^{0}+\phi^{0}$ will be smaller than the universe expansion rate when $T\lesssim10$ GeV, refer to \cite{23}, thus this portal is closed. The other one is the normal neutrino coupling in Eq. (11), by which the SM $\nu_{L}^{0}$ is connected to the dark sector. However, at $T_{D}\approx1$ MeV, $\nu_{L}^{0}$ also decouples from the SM sector and disappears into the dark sector, it thus becomes the last member of the dark sector. From that time on, the dark sector and the SM one are isolated from each other.

  Below $T_{D}\approx1$ MeV, the SM sector and the dark sector separately evolve without communications, therefore the entropy in each sector is respectively conserved, we can therefore derive the $\nu^{0}$ effective temperature such as
\begin{alignat}{1}
 &\frac{s^{\mathrm{DS}}(T_{D})a^{3}(T_{D})}{s^{\mathrm{SM}}(T_{D})a^{3}(T_{D})}
  =\frac{s^{\mathrm{DS}}(T_{0})a^{3}(T_{0})}{s^{\mathrm{SM}}(T_{0})a^{3}(T_{0})}\,,\nonumber\\
 \Longrightarrow&\frac{g_{*}^{\mathrm{DS}}(T_{D})}{g_{*}^{\mathrm{SM}}(T_{D})}=\frac{g_{*}^{\nu}+g_{*}^{N}+g_{*}^{\phi^{0}}}{g_{*}^{e}+g_{*}^{\gamma}}
  =\frac{g_{*}^{\nu}}{g_{*}^{\gamma}}(\frac{T_{\nu}}{T_{0}})^{3},\nonumber\\
 \Longrightarrow&\frac{T_{\nu}}{T_{0}}=(\frac{16}{21})^{\frac{1}{3}},\hspace{0.3cm} T_{\nu}\approx2.49\:\mathrm{K},
\end{alignat}
  where $a(T)$ is the scale factor of the universe expansion. At $T_{D}\approx1$ MeV, the $Z_{2}^{M}$ symmetry is yet unbroken, so the dark particles are all massless states. At $T=v_{0}\approx0.1$ MeV, the $Z_{2}^{M}$ parity violation gives rise to the light masses of the dark particles, among which only $\nu^{0}$ is still relativistic state. Here $T_{\nu}\approx2.49$ K is higher than $Y_{\nu}\approx1.95$ K given by the SM, this a prediction of the model.

  On the basis of the last two terms of the third line in Eq. (11), the main evolutions inside the dark sector are the following processes,
\begin{alignat}{1}
 &\phi^{0}\rightarrow N_{1}+\overline{N}_{1},\hspace{0.3cm} N_{2,3}+\overline{N}_{2,3}\rightarrow N_{1}+\overline{N}_{1},\hspace{0.3cm}
  N_{1}+\overline{N}_{1}\rightarrow \nu+\overline{\nu},\nonumber\\
 &N_{1}+N_{1}\rightarrow N_{1}+N_{1}\,,\hspace{0.3cm} N_{1}+\overline{N}_{1}\rightarrow N_{1}+\overline{N}_{1}\,,\hspace{0.3cm}
  N_{1}+\nu\rightarrow N_{1}+\nu\,.
\end{alignat}
  $\phi^{0}$ can decay into $N_{1}+\overline{N}_{1}$ only if $m_{\phi^{0}}>2m_{N_{1}}$, so it is absence in the present-day universe. However, $N_{i}$ is stable without decay, its only way out is therefore $N_{i}+\overline{N}_{i}$ annihilating into a pair of lighter particles via the s-channel $\phi^{0}$ mediation, Fig. 3 shows the relevant Feynman diagrams.
\begin{figure}
 \centering
 \includegraphics[totalheight=4cm]{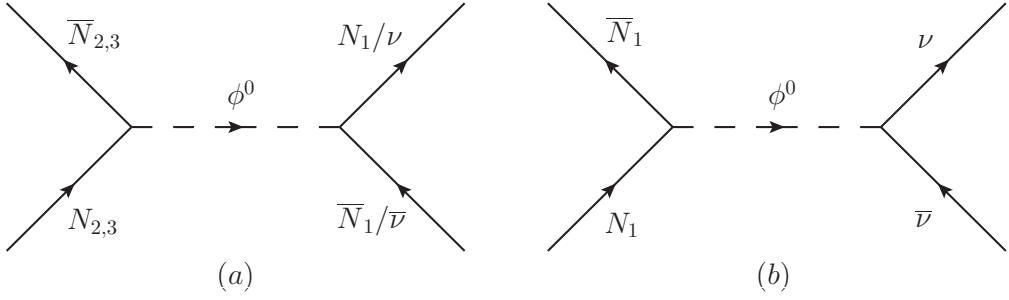}
 \caption{(a) $N_{2,3}+\overline{N}_{2,3}\rightarrow N_{1}+\overline{N}_{1}$ has a very strong cross-section, so the heavier pairs of $N_{2,3}+\overline{N}_{2,3}$ are wholly annihilating exhaustion. (b) $N_{1}+\overline{N}_{1}\rightarrow\nu+\overline{\nu}$ has a weak cross-section, which exactly fits the ``WIMP Miracle", so the lightest pair of $N_{1}+\overline{N}_{1}$ can remain sizeable relics. Below the freeze-out temperature of $T_{f}\sim1$ KeV, $N_{1}$ and $\overline{N}_{1}$ are non-relativistic decoupling and become the CDM, at the same time, $\nu$ and $\overline{\nu}$ are relativistic decoupling and become the HDM.}
\end{figure}
  Because of $Y_{N}\sim1\gg Y_{\nu}\sim10^{-6}$, the annihilation cross-section of $N_{2,3}+\overline{N}_{2,3}\rightarrow N_{1}+\overline{N}_{1}$ is $\sim10^{10}$ times larger than one of $N_{i}+\overline{N}_{i}\rightarrow\nu+\overline{\nu}$, namely, the former is a very strong annihilation, whereas the latter has only a weak cross-section. As a result, the heavier pairs of $N_{2,3}+\overline{N}_{2,3}$ are wholly annihilating exhaustion so that they are absence in the present-day dark sector, by contrast, the lightest pair of $N_{1}+\overline{N}_{1}$ can not completely be annihilating exhaustion, thus sizeable relics of them are left in the dark sector, which are namely the CDM in the present-day universe. The second line in Eq. (21) are all elastic scatterings via the $\phi^{0}$ mediation, whose implications will be explained in the end of this Section.

  When the temperature decreases to the freeze-out temperature, the annihilate rate of $N_{1}+\overline{N}_{1}\rightarrow\nu+\overline{\nu}$ is smaller than the universe expansion rate, thus the annihilation process is closed and their relic density is frozen in the dark sector. As a result, $N_{1}$ and $\overline{N}_{1}$ are non-relativistic decoupling and become the CDM, at the same time, $\nu$ and $\overline{\nu}$ are relativistic decoupling and become the HDM. The annihilate cross-section and the freeze-out temperature are calculated as follows,
\begin{alignat}{1}
 &\Gamma[N_{1}+\overline{N}_{1}\rightarrow\nu+\overline{\nu}]=\langle\sigma v_{r}\rangle n_{N_{1}}=H(T_{f}),\nonumber\\
 &n_{N_{1}}(T_{f})=2T_{f}^{3}(\frac{m_{N_{1}}}{2\pi T_{f}})^{\frac{3}{2}}e^{-\frac{m_{N_{1}}}{T_{f}}},\nonumber\\
 &\langle\sigma v_{r}\rangle_{T_{f}}=a+b\,\langle v_{r}^{2}\rangle_{T_{f}}=a+b\,\frac{6T_{f}}{m_{N_{1}}}\,,\nonumber\\
 &a=0,\hspace{0.3cm} b=\frac{\sum\limits_{i}m^{2}_{\nu_{i}}}{128\pi v_{0}^{4}(1-y)^{2}}\,,\hspace{0.3cm}
  y=(\frac{m_{\phi^{0}}}{2m_{N_{1}}})^{2},\nonumber\\
 \Longrightarrow &\frac{m_{N_{1}}}{T_{f}}=\frac{1}{x}\approx 11.4+ln\frac{m_{N_{1}}(\mathrm{MeV})}{\sqrt{xg_{*}(T_{f})}}
  +ln\frac{\langle\sigma v_{r}\rangle_{T_{f}}(\mathrm{GeV}^{-2})}{10^{-10}}\,,\nonumber\\
 &\mathrm{for}\;v_{0}\sim0.1\:\mathrm{MeV},\;y\sim10,\;m_{N_{1}}\sim0.01\:\mathrm{MeV},\nonumber\\
 \Longrightarrow &\langle\sigma v_{r}\rangle_{T_{f}}\sim5\times10^{-9}\:\mathrm{GeV^{-2}},\hspace{0.3cm} \frac{m_{N_{1}}}{T_{f}}\sim10,
\end{alignat}
  where $v_{r}=2\sqrt{1-\frac{4m^{2}_{N_{1}}}{s}}$ is a relative velocity of $N_{1}$ and $\overline{N}_{1}$, $\langle\sigma v_{r}\rangle_{T_{f}}$ is the thermally averaged annihilate cross-section, note that the s-wave contribution to it is vanishing, namely $a=0$. For these parameter values in Eq. (22), $\langle\sigma v_{r}\rangle_{T_{f}}$ is exactly a weak interaction cross-section, which is namely the so-called ``WIMP Miracle" \cite{24}, and then we can derive the freeze-out temperature of $T_{f}\sim1$ KeV and $g_{*}(T_{f})=g_{*}^{\gamma}+g_{*}^{\nu}(\frac{T_{\nu}}{T_{0}})^{3}=10$.

  In the present-day universe, the density of the CDM consisting of $N_{1}$ and $\overline{N}_{1}$ is calculated by the following equations \cite{25},
\begin{alignat}{1}
 &\Omega_{\mathrm{CDM}}h^{2}=\Omega_{N_{1}+\overline{N}_{1}}h^{2}=\frac{2m_{N_{1}}n_{N_{1}}(T_{0})}{\rho_{c}}\,h^{2}
  =\frac{0.87\times10^{-10}\:\mathrm{GeV}^{-2}}{\sqrt{g_{*}(T_{f})}\,x(a+3bx)}\approx0.119,\nonumber\\
 &n_{N_{1}}(T_{0})=\frac{g_{*}(T_{0})T_{0}^{3}}{g_{*}(T_{f})T_{f}^{3}}\,n_{N_{1}}(T_{f})\,.
\end{alignat}
  By use of $a,b,x$ given in Eq. (22), we can correctly reproduce $\Omega_{\mathrm{CDM}}h^{2}\approx0.119$, which is the current density of the CDM from the multiple observations \cite{1,26}. On the other hand, the density of $\nu$ and $\overline{\nu}$ as the HDM is given by the following relations,
\begin{alignat}{1}
 &\Omega_{\mathrm{HDM}}h^{2}=\Omega_{\nu+\overline{\nu}}h^{2}=\frac{n_{\nu}(T_{\nu})\sum\limits_{i}m_{\nu_{i}}}{\rho_{c}}\,h^{2}\approx3.5\times10^{-3},\nonumber\\
 &n_{\nu}(T_{\nu})=(\frac{T_{\nu}}{T_{f}})^{3}n_{\nu}(T_{f})=\frac{1.2}{\pi^{2}}\,g'_{\nu}T_{\nu}^{3}\approx469\:\mathrm{cm}^{-3},
\end{alignat}
  where $g'_{\nu}=\frac{3}{4}\times4=3$ for one generation of massive Dirac neutrino. Here the neutrino number density $n_{\nu}$ is about four times as large as $n_{\nu}\approx112\:\text{cm}^{-3}$ given by the SM, moreover, it exceeds $n_{\gamma}(T_{0})\approx411\:\text{cm}^{-3}$. The above density value of the HDM neutrino is another prediction of the model, see the following Eq. (26). However, both the CDM $N_{1}$ and the HDM $\nu$ are in the dark sector, they are isolated from the SM sector at the low energy, so it is very difficult to detect them.

  Finally, we explain implications of those elastic scatterings of the second line in Eq. (21). In fact, $N_{1}+N_{1}\rightarrow N_{1}+N_{1}$ and $N_{1}+\overline{N}_{1}\rightarrow N_{1}+\overline{N}_{1}$ imply a self-interaction among the CDM via the $\phi^{0}$ mediation, these elastic scatterings can still in equilibrium after the weak annihilation of $N_{1}+\overline{N}_{1}\rightarrow\nu+\overline{\nu}$ was frozen, therefore this self-interaction can drive the distribution of the CDM with the frozen density and impact on the structure formation, in particular, has effect on small scale structure of the universe \cite{27}, we will specially discuss this problem in another paper. In addition, the elastic scattering of $N_{1}+\nu\rightarrow N_{1}+\nu$ means a weak interaction between the CDM and the HDM, of course, it is also frozen at $T_{f}\sim1$ KeV, so the CDM and the HDM are also isolated from each other in the present universe. In conclusion, the model not only completely explains the origin of the dark matter, but also sheds light on detections of the attractive dark universe.

\vspace{0.6cm}
\noindent\textbf{V. Numerical Results and Model Test}

\vspace{0.3cm}
  We now demonstrate and summarize the model by some concrete numerical results. All kinds of the parameters in the SM sector have essentially been fixed by the current experimental data \cite{1}. Some key parameters in the mirror sector, also including the dark sector, can be determined jointly by the current data of the tiny neutrino mass, the baryon asymmetry, and the CDM abundance. Based on a whole fit, the key parameters of the model are therefore chosen as follows,
\begin{alignat}{1}
 &v_{L}=246\:\mathrm{GeV},\hspace{0.3cm} v_{R}=10^{6}\:\mathrm{GeV},\nonumber\\
 &\mathrm{Tr}(Y_{L}^{\dagger}Y_{L})=\mathrm{Tr}(Y_{R}^{\dagger}Y_{R})=\mathrm{Tr}(Y_{3}^{\dagger}Y_{3})=1,\hspace{0.3cm}
  (Y_{R}^{\dagger}Y_{R})_{11}=(Y_{3}^{\dagger}Y_{3})_{11}=10^{-6},\nonumber\\
 &M_{\phi^{-}}=2\times10^{4}\,(5\times10^{4})\:\mathrm{GeV},\hspace{0.3cm}
  v_{0}=0.05\,(0.1)\:\mathrm{MeV},\hspace{0.3cm} m_{N_{1}}=0.1v_{0}\,,\nonumber\\
 &\frac{M_{\chi_{1}}}{M_{\phi^{-}}}=0.135\,(0.125),\hspace{0.2cm} \frac{M_{\chi_{2}}}{M_{\phi^{-}}}=0.176\,(0.19),\hspace{0.2cm}
  \frac{M_{\chi_{3}}}{M_{\phi^{-}}}=3.62\,(2.9),\hspace{0.2cm} \frac{m_{\phi^{0}}}{2m_{N_{1}}}=3.78\,(1.98),
\end{alignat}
 where those values in the first two lines are fixed as benchmark, the last two lines are two sets of typical values in the parameter space (the second set are inside brackets). For the two sets of values of $M_{\phi^{-}}$ and $v_{0}$, firstly we can determine $\frac{M_{\chi_{3}}}{M_{\phi^{-}}}$ and $\frac{M_{\chi_{2}}}{M_{\phi^{-}}}$ by fitting the two mass-squared differences of the neutrino, secondly $\frac{M_{\chi_{1}}}{M_{\phi^{-}}}$ is determined by fitting the baryon asymmetry $\eta_{B}$, lastly the ratio of $\frac{m_{\phi^{0}}}{2m_{N_{1}}}$ is determined by the CDM density $\Omega_{\mathrm{CDM}}h^{2}$. It can be seen from Eq. (25) that $m_{N_{1}}$ is in the range of $5-10$ KeV, while $M_{\chi_{1}}$ is about $3-6$ GeV. In short, all of the parameter values in Eq. (25) are consistent and reasonable, moreover, without any fine-tuning, they are completely in accordance with the model requirements and the previous discussions, see Eq. (14).

  Now we substitute Eq. (25) into the previous relevant equations of the model, then we correctly reproduce the following desired results,
\begin{alignat}{1}
 &m_{\nu_{1}}=0.0155\,(0.0119)\:\mathrm{eV},\hspace{0.3cm} m_{\nu_{2}}=0.0178\,(0.0147)\:\mathrm{eV},\hspace{0.3cm}
  m_{\nu_{3}}=0.0534\,(0.0522)\:\mathrm{eV},\nonumber\\
 &\triangle m_{21}\approx 7.52\,(7.58)\times10^{-5}\:\mathrm{eV^{2}},\hspace{0.3cm}
  \triangle m_{32}\approx 2.54\,(2.51)\times10^{-3}\:\mathrm{eV^{2}},\nonumber\\
 &\frac{\Gamma}{H}=0.427\,(0.136),\hspace{0.3cm} \eta_{B}\approx6.12\,(6.07)\times10^{-10},\hspace{0.3cm}
  \Omega_{B}h^{2}\approx0.0224\,(0.0222),\nonumber\\
 &\Omega_{\mathrm{CDM}}h^{2}\approx0.119\,(0.119),\hspace{0.3cm} \Omega_{\mathrm{HDM}}h^{2}\approx0.00386\,(0.00351),
\end{alignat}
  where $\triangle m_{ij}=m^{2}_{\nu_{i}}-m^{2}_{\nu_{j}}$, and $\frac{\Gamma}{H}$ is the ratio of $\Gamma[\chi_{1}^{-}\rightarrow\nu_{R}^{0c}+l_{L}+l_{L}]$ to $H(M_{\chi_{1}})$. These values of $\frac{\Gamma}{H}$ are all smaller than one, this thus confirms that the decay is indeed out-of-equilibrium which is a necessary prerequisite. Explicitly, all the results of Eq. (26) are very well in agreement with the current experimental data \cite{1}. In conclusion, only by use of these simple and natural parameters in Eq. (25), the model can completely and satisfactorily account for the three outstanding puzzles of the neutrino mass, the baryon asymmetry, and the dark matter, so this sufficiently demonstrates that the model is very successful and believable.

  In the end, any particle theory has to be tested by experiments, here we simply discuss several approaches to test the model. The heavy mirror particles can not be produced at the present colliders, but we can search their decay products through high-energy cosmic rays, for example, the searches for $\chi^{-}\rightarrow e^{-}+\phi^{0}$ and $Z_{R}^{0}\rightarrow e^{-}+e^{+}$ or $\nu_{L}^{0}+\nu_{L}^{0c}$. The latest news from DAMPE collaboration about the cosmic ray spectrum from 40 GeV to 100 TeV, refer to \cite{28}, may provide an opportunity for such searches. On the basis of those couplings of the model, we can also test the model predictions and probe the dark sector by the three feasible approaches shown as Fig. 4.
\begin{figure}
 \centering
 \includegraphics[totalheight=4cm]{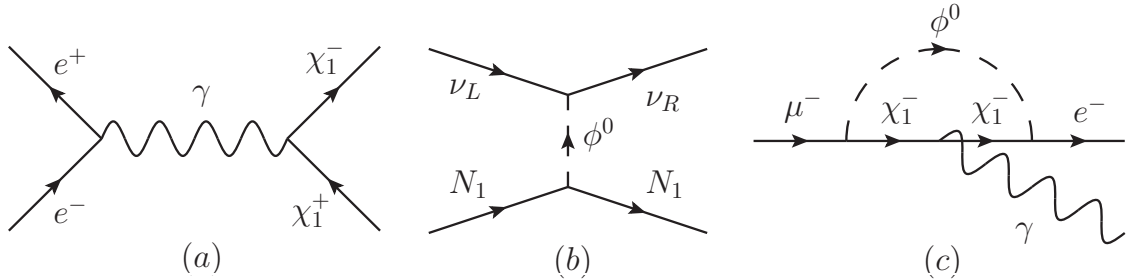}
 \caption{(a) The pair production of the lightest mirror charged lepton at the future $e^{-}+e^{+}$ collider with $\sqrt{s}=10$ TeV, then the $CP$-asymmetric decay of $\chi_{1}^{\pm}$ can generate the SM lepton asymmetry. (b) The elastic scattering of the SM $\nu_{L}$ and the CDM $N_{1}$, by which $\nu_{L}$ is converted into the dark $\nu_{R}$ so that it escapes the detector. (c) The LFV process of $\mu^{-}\rightarrow e^{-}+\gamma$ through both $\chi_{1}^{-}$ and $\phi^{0}$ mediation.}
\end{figure}

  The (a) diagram shows that the lightest mirror charged lepton pair is directly produced by the future $e^{-}+e^{+}$ collider with $\sqrt{s}=10$ TeV such as ILC \cite{29}, then the $CP$-asymmetric decay of $\chi_{1}^{\pm}$ can further generate an asymmetric number of the SM lepton and anti-lepton in the final states. Therefore this method can directly test the leptogenesis mechanism in the model.

  The (b) diagram shows the elastic scattering of $\nu_{L}+N_{1}\rightarrow\nu_{R}+N_{1}$ via the t-channel $\phi^{0}$ mediation, note that the $\nu $ chirality is changed in this process. If a beam of the SM $\alpha$-flavor $\nu_{L\alpha}$ is produced at the laboratory, on its way to the distant detector, its tiny part can be scattered by the surrounding CDM $N_{1}$ and converted into the dark right-handed $\nu_{R}$, thus they escape the detector, so we can detect the model dark sector by use of this method. The scattering cross-section is given by
\begin{alignat}{1}
 &\sum\limits_{\beta}\sigma[\nu_{L\alpha}+N_{1}\rightarrow \nu_{R\beta}+N_{1}]=
  \frac{(M_{\nu}^{\dagger}M_{\nu})_{\alpha\alpha}}{64\pi v^{4}_{0}E^{2}_{\nu_{L\alpha}}}f(E_{\nu_{L\alpha}}),\nonumber\\
 &M_{\nu}^{\dagger}M_{\nu}=U_{\nu_{L}}\,\mathrm{Diag}(m^{2}_{\nu_{1}},m^{2}_{\nu_{2}},m^{2}_{\nu_{3}})U_{\nu_{L}}^{\dagger},\nonumber\\
 &f(E_{\nu_{L\alpha}})=\int\limits^{t_{0}}\limits_{t_{1}}dt\frac{t(t-4m^{2}_{N_{1}})}{(t-m^{2}_{\phi^{0}})^{2}}\,,\hspace{0.3cm}
  t_{0}=0,\hspace{0.3cm} t_{1}=-\frac{4E^{2}_{\nu_{L\alpha}}m_{N_{1}}}{2E_{\nu_{L\alpha}}+m_{N_{1}}}\,,
\end{alignat}
  where $E_{\nu_{L\alpha}}$ is the $\nu_{L\alpha}$ energy in the laboratory frame and $U_{\nu_{L}}$ is the $\nu_{L\alpha}$ mixing matrix measured by the experiments. If we can use the electronic neutrino beam with $E_{\nu_{Le}}=1$ MeV, then $f(E_{\nu_{Le}})\approx-t_{1}$, thus we can estimate $\sigma\sim10^{-10}$ $\text{GeV}^{-2}$ provided $v_{0}=0.1$ MeV and $m_{N_{1}}=0.1v_{0}$, which is also a weak interaction cross-section, similar to one in Eq. (22). Besides this scattering cross-section, the scattering rate depends on the local density of the CDM $N_{1}$, therefore we can learn the information of the dark physics such as $v_{0}$ and $m_{N_{1}}$ by means of measuring the $\nu_{Le}$ disappearance rate. In fact, the cosmic neutrino source is a better laboratory, for instance, we can detect the $\nu_{L}$ stream emitted by a distant supernova, it will travel through the CDM in the galactic halo before it can arrive to the earth, its tiny part will be scattering off and converted into the dark $\nu_{R}$, thus the $\nu_{L}$ stream which is eventually received is certainly less than the expected value. This detection is very similar to one of the flavor oscillation of the solar neutrino. In a word, this method can not only detect the dark sector physics, but also corroborate the neutrino mass origin in the model.

  The (c) diagram is a LFV process of $\mu\rightarrow e\gamma$ through both $\chi_{1}^{-}$ and $\phi^{0}$ mediation. Its branch ratio is estimated as $\sim10^{-14}$ provided $(Y_{3}^{\dagger})_{12}(Y_{3})_{11}\sim10^{-4}$, which is one order of magnitude lower than the current limit \cite{1}, so this process is very promising to be detected in the near future.

  In short, the above suggestions can be considered as new subjects and goals of the experimental physicists which are endeavoring to search new physics evidences beyond the SM \cite{5,30}. Although it will be very large challenges to actualize them, it is not impossible, moreover, its scientific significance is beyond all doubt.

\vspace{0.6cm}
\noindent\textbf{VI. Conclusions}

\vspace{0.3cm}
  In summary, I suggest the left-right mirror symmetric particle model as the natural and aesthetic extension of the SM. This model has the left-right symmetric gauge group of $SU(2)_{L}\times U(1)_{Y}\times SU(2)_{R}$, and also it conserves the global $B-L$ number and the discrete $Z_{2}^{M}$ matter parity. At the $\sim10^{6}$ GeV scale the $SU(2)_{R}$ breaking gives rise to the heavy mirror particle masses, at the $\sim0.1$ MeV scale the $Z_{2}^{M}$ violating generates the light dark particle masses, the SM electroweak breaking lies just between the two scale, but the $B-L$ number is always conserved. The tiny neutrino mass results from the weak effective Dirac neutrino coupling which is generated by the loop diagram radiation. The $CP$-asymmetric and out-of-equilibrium three-body decay of the lightest mirror charged lepton can lead to the $B-L$ asymmetry in SM sector and the $\nu_{R}$ asymmetry, the former is partly converted into the baryon asymmetry through the sphaleron effect, the latter disappears into the dark sector due to the $\nu_{R}$ decoupling. The dark sector consists of all of the light neutral particles except the photon, note that $\nu_{L}$ eventually disappears into the dark sector after it decoupling from the SM sector. The dark Dirac fermion $N_{1}$ with $\sim10$ KeV mass is a desirable CDM candidate. $N_{1}+\overline{N}_{1}$ can annihilate into $\nu+\overline{\nu}$ via the dark scalar $\phi^{0}$ mediating, the annihilation cross-section exactly fits the ``WIMP Miracle". Below the freeze-out temperature of $\sim1$ KeV, $N_{1}$ and $\nu$ are decoupling from each other, respectively, become the CDM and the HDM in the present universe.

  In short, the model can simply and completely account for the common origin of the tiny neutrino mass, the baryon asymmetry and the dark matter, moreover, profoundly uncover the internal connections among them. In addition, the model gives some interesting predictions, for instance, the lightest mirror charged lepton mass is about several TeVs, the dark physics scale is $\sim0.1$ MeV, the $\nu$ effective temperature is $2.49$ K, the HDM $\nu$ density is $\sim3.5\times10^{-3}$, and so on. Finally, I give several feasible approaches to test the model by means of the TeV line collider, the neutrino experiments, the detection for $\mu\rightarrow e\gamma$, and the search for high-energy cosmic rays. The fruitful mirror and dark physics world are waiting for us to explore. In the near future, it is very possible that we shall ushered in a new physics era beyond the SM and open the door of the dark universe.

\vspace{0.6cm}
 \noindent\textbf{Acknowledgements}

\vspace{0.3cm}
  I would like to thank my wife for her great helps. This research is supported by the Fundamental Research Funds for the Central Universities of China under Grant No. WY2030040065.

\end{document}